\documentclass[10pt,a4paper]{article}
\usepackage[latin1]{inputenc}
\usepackage{amsmath}
\usepackage{amsfonts}
\usepackage{amssymb}
\usepackage{graphics}
\usepackage{graphicx}
\usepackage{subfigure}
\begin{document}
\setlength{\unitlength}{1.0cm}

\author{D. Yadykin, Y.Q. Liu, R. Paccagnella}
\title{{\bf Effect of kinetic resonances on the stability of Resistive Wall Mode in Reversed Field Pinch}}
\maketitle
\section{Introduction}
Understanding  physics and stabilization of the Resistive Wall Mode (RWM) is an important task for the successful operation of the present day and the future fusion devices. Converted from the ideal MHD mode in the presence of the wall with finite conductivity, and growing on the time scale of the magnetic field penetration time through the wall, RWM sets pressure limit in the advanced scenario of the tokamak device \cite{garofallo_prl_99_1, gryaznev_baps_03_1, shilov_pop_04_1}. It also limits the discharge duration of the Reversed Field Pinch (RFP) device \cite{alper_ppcf_89_1, brunsell_pop_03_1, bolzonella_32eps_1}. Control of RWM growth is necessary  for the discharge times longer than the wall time. 
 
 Two general methods are known for the RWM stabilization: active control and stabilization via the mode resonance with continuum spectra or particle motions. For the active control the perturbed magnetic field measured by the set of sensors facing the plasma surface is used to generate the control signal by the set of active control coils. Such mechanism is extensively studied theoretically \cite{bishop_ppcf_89_1, fitzpatrick_pop_96_1, okabayashi_nf_98_1, fitzpatrick_pop_99_1, liu_pop_00_1,pustovitov_ppr_01_1} and is successfully applied in the present day fusion devices, allowing operation with the plasma pressures above no-wall limit for tokamaks \cite{okabayashi_pop_01_1,strait_pop_04_1,sabbagh_prl_04_1} and resulting in substantial increase of the plasma discharge duration in RFPs\cite{brunsell_prl_04_1, pacaggnella_prl_06_1} .
  
  RWM suppression via the mode resonance with continuum spectra or particle motions is another possible control mechanism (also called 'rotational stabilization'). It was experimentally observed for the tokamak configuration \cite{reimerdes_phys_plasams_06_1} that RWM is stabilized when the plasma rotation frequency is sufficiently high. Several mechanisms were proposed that could explain the stabilization. In ideal MHD description the dissipation caused by the RWM resonance with Alfv\'{e}n \cite{gregoratto_ppcf_01_1, zheng_prl_05_1} or sound \cite{bondeson_prl_94_1, betti_prl_95_1} continuum spectra could lead to the RWM stabilization. The dissipation strength depends on the plasma rotation frequency. The plasma rotation frequency $\Omega$ of the order of sound frequency $\omega_s$ is needed for the RWM stabilization by the resonance with continuum spectrum. Another mechanism that could lead to the RWM suppression in the region $\Omega\sim\omega_s$ is ion Landau damping. Accurate description of this mechanism requires kinetic treatment of the ion motion parallel to the magnetic field \cite{bondeson_ph_fluids_89_1}. In ideal MHD description this mechanism is modelled by adding dissipation term \cite {chu_phys_plasmas_95_1}. More physically consistent approach is used in semi-kinetic model \cite{bondeson_phys_plasmas_96_1} where damping term is calculated using drift kinetic energy principle for the large aspect ration approximation. Mode resonance with circulating and trapped ions are included in the model. Calculated damping term are then included in the MHD description. In order to explain observed RWM stabilization for the very low plasma rotation frequency \cite{reimerdes_prl_07_1} ($\Omega \ll \omega_s$) other model was proposed \cite{hu_prl_04_1} where the main dissipation channel is due to the mode resonance with the precession drift motion of the trapped particles. This model in fact predicts the RWM stabilization even without plasma rotation. The accurate prediction of the RWM stabilization for the low plasma rotation frequencies is an important issue for the future reactor experiment ITER, where the highest plasma rotation in steady-state scenario is predicted to be in the sound range \cite{polevoi_02_1}.

Previous studies of the rotational stabilization of the RWM  were performed mainly for the tokamak configuration. Available results for the RFP \cite{jiang_phys_plasmas_95_1, guo_phys_plasmas_99_1} show that the plasma rotation frequency in the range of the Alfv\'{e}n frequency ($\Omega \sim \omega_a$) is needed for the RWM stabilization by the mode resonance with continuum spectrum. Such high plasma rotation is not observed in the present day RFP devices, therefore rotational stabilization mechanism was not considered as realistic for RFP. On the other hand mode resonance with particle motions occurs for the low plasma rotation frequency and could be considered as a possible RWM stabilization mechanism also for RFP configuration. In this work numerical studies of the effect of kinetic resonances on the RWM stability are performed for reversed-field pinch (RFP) configuration. Particularly, parameters corresponding to the RFX device \cite{sonato_fid_03_1} is used. In RFP the values of poloidal and toroidal components are of the same order of magnitude. Toroidal component of the equilibrium magnetic field changes sign at the plasma edge. Several Fourier harmonics with different poloidal and toroidal mode numbers are seen in the RFP mode spectrum. Resonant harmonics with poloidal mode numbers $m=0,1$ are usually linearly stable. Several non-resonant harmonics with different toroidal mode numbers (positive and negative according to the sign of the safety factor $q$ of the resonant magnetic surface) are unstable growing on the times comparable with the equilibrium magnetic field penetration through the resistive wall and therefore are classified as RWMs. Experimental studies of RWM control in RFP have shown possibility of simultaneous suppression of several unstable harmonics with different toroidal mode numbers using active feedback \cite{brunsell_prl_04_1, pacaggnella_prl_06_1}. 
  
\section{Model} \label{sec2}
In this section physical model is briefly described that is used in the present work. More complete derivation can be found in \cite{liu_phys_plasmas_08_1}. Stability of RWM for these studies is determined by solving numerically the system of MHD equations with  toroidal plasma rotation:
\begin{align}
(-i\omega+in\Omega)\mathbf{\xi}&=\mathbf{v}+(\mathbf{\xi} \cdot \nabla\Omega)R^2\nabla \phi \label{eq1}\\
\rho(-i\omega+in\Omega)\mathbf{v}&=-\nabla \cdot \mathbf{p} +\mathbf{j}\times \mathbf{B}+\mathbf{J}\times \mathbf{Q}- \rho[2\Omega\mathbf{\hat{Z}}\times \mathbf{v} +(\mathbf{v} \cdot\nabla\Omega)R^2 \nabla\phi] \label{eq2}\\
(-i\omega+in\Omega)\mathbf{Q}&=\nabla\times(\mathbf{v}\times \mathbf{B})+(\mathbf{Q}\cdot\nabla\Omega)R^2\nabla\phi \label{eq3}\\
(-i\omega+in\Omega)p&=-\mathbf{v}\cdot\nabla P \label{eq4}\\
\mathbf{j}&=\nabla\times \mathbf{Q} \label{eq5}
\end{align}
where $\omega$ is the complex mode frequency ($\omega=i\gamma-\omega_r$ where $\gamma$ - mode growth or damping rate and $\omega_r$ - mode rotation frequency), $\mathbf{B}$,$\mathbf{J}$,$P$- equilibrium magnetic field, current density and pressure respectively, $\mathbf{\hat{Z}}$ - unit vector in the vertical direction, $\rho$ - plasma density, $\Omega$ - plasma rotation frequency in toroidal direction $\phi$, $\mathbf{\xi}$,$\mathbf{v}$,$\mathbf{j}$,$\mathbf{Q}$ - plasma displacement,perturbed velocity,perturbed current,perturbed magnetic field respectively, $\textbf{p}$ - pressure tensor.  

Kinetic terms are included into the MHD equations via the pressure tensor components. Pressure tensor is defined as 
\begin{equation}
\mathbf{p}=\textbf{I}p+p_\parallel\mathbf{\hat{b}\hat{b}}+p_\bot (\mathbf{I}-\mathbf{\hat{b}\hat{b}})
\label{pres_tensor}
\end{equation} 
where $p$ is the scalar fluid pressure perturbation, $p_\parallel,p_\bot$ are the particle parallel and perpendicular components of the kinetic pressure perturbations, $\mathbf{\hat{b}}=\textbf{B}/B$, $B=\mid \textbf{B} \mid$, $\textbf{I}$ is the unit tensor. For the particular perturbation with the mode number $n$, the parallel and perpendicular components of the pressure tensor are written as:
\begin{align}
 p_\parallel e^{-i\omega t+in\phi}& = \sum_{e,i}\int\int d\epsilon d\Lambda M v_\parallel^2 f_L^1 \\
 p_\bot e^{-i\omega t+in\phi}& = \sum_{e,i}\int\int d\epsilon d\Lambda \frac{1}{2} M v_\bot^2 f_L^1
 \end{align}
 Here the summation is over the electron and ion plasma components, integration is carried out over the particle energy $\epsilon$ and pitch angle $\Lambda$, M is the particle mass, $v_\parallel, v_\bot$ are the parallel and perpendicular velocity components respectively with respect to the equilibrium magnetic field, $f_L^1$ is the perturbed particle distribution function. It is derived as:
 \begin{equation}
f_L^1=-f_\epsilon^0\epsilon_k e^{-i\omega t+in\phi}\sum_{m,l} X_m H_{ml}\lambda_{ml}e^{-in\tilde{\phi}(t)+im<\dot{\chi}>+il\omega_b t}
\label{pert_distr}
 \end{equation}
where subscripts $n$,$m$,$l$ mark Fourier components along toroidal and polidal angles and along bounce particle orbit, $f_\epsilon^0$ - derivative of the equilibrium distribution function (taken to be Maxwellian for thermal particles) with respect to the particle energy $\epsilon$, $\epsilon_k$ - kinetic energy of the particles, $H_{ml}$ - perturbed Lagrangian component in Fourier space, $\lambda_{ml}$ mode-particle resonance operator, $\tilde{\phi}(t)=\phi(t)-<\dot{\phi}> t $ - the periodic component of the particle velocity in toroidal direction, $<..>$ means average over the particle bounce period. The mode-particle resonant operator is calculated as:
 \begin{equation}
     \lambda_{ml}=\frac{n(\omega_{*N}+(\epsilon_k-3/2)\omega_{*T}+\Omega)-\omega}{n\omega_d+ n\Omega+[\alpha(m+nq)+l]\omega_b-i\nu_{eff} -\omega}
     \label{res_oper}
     \end{equation}
 where  $\omega_{*N},\omega_{*T}$ - diamagnetic frequencies associated with the density and temperature gradients respectively, $\omega_b$ - bounce frequency, $\omega_d$ - bounce averaged magnetic drift frequency, $\nu_{eff}$ - effective collision frequency, $\alpha=$0 for the trapped particles, $\alpha=$1 for the passing particles. The energy transfer between the mode and the particles is described by the imaginary part of the resonant operator. It can be seen that in case of finite mode growth (damping) rate and collision frequency energy is transferred between the mode and the particles for all plasma rotation frequency values. Mode-particle interaction is enhanced in the resonant regions, where the real part of the resonant operator is equal to zero. The collisionallity effect on the RWM stability (via collision frequency term $i\nu_{eff}$ of the resonant operator) is not considered in this work.
     

Using \eqref{pert_distr} and \eqref{res_oper} to calculate the pressure tensor components and substituting \eqref{pres_tensor} in \eqref{eq2}, the self-consistent formulation is obtained, that allows to study RWM stability including the effect from kinetic resonances. 

System of ideal MHD equations where the ion Landau damping is modelled by the parallel viscosity can be obtained from the equations \eqref{eq1} - \eqref{eq5} by using only scalar pressure part of the pressure tensor \eqref{pres_tensor}, adding the viscous stress tensor $\nabla \cdot \Pi=\kappa \vert k_\Vert v_{thi}\vert\rho(\mathbf{v \cdot \hat{b}\hat{b}})$ in eq. \eqref{eq1}, where $\kappa$ is the coefficient determining the damping strength, $k_\Vert=(n-\frac{m}{q})/R$, $v_{thi}$ is the ion thermal velocity and adding the term $-5/3P \nabla \cdot \mathbf{v}$ in the eq. \eqref{eq4}. 

\section{Numerical details}

System of equations \eqref{eq1} - \eqref{eq5} is an eigenvalue problem that can be written in the matrix form as
\begin{equation}
 \gamma_0 B X=A X
 \label{eigen}
\end{equation}
where $\gamma_0$ is the complex eigenvalue, $A,B$ - MHD operators, $X$ - eigenvector. In this work the eigenvalue problem \eqref{eigen} is solved numerically using MHD stability code MARS-K \cite{liu_phys_plasmas_08_1} in toroidal coordinates ($s$,$\chi$,$\phi$), where $s=\sqrt{1-\frac{\psi}{\psi_0}}$ is the normalized poloidal flux ($\psi_0$ is the poloidal flux value on the axis), $\chi$ is the generalized poloidal angle and $\phi$ is the geometrical toroidal angle. Equilibrium quantities are obtained using equilibrium solver CHEASE \cite{lutjens_cpc_96_1}.  Note that eigenvalue problem is non-linear as resonant operator \eqref{res_oper} includes eigenvalue $\gamma_0$ and therefore $A=A(\gamma_0)$. Outer iterative loop is required in order to obtain the solution of the eigenvalue problem. 
 
\section{Results}
   \subsection{Equilibrium} \label{eqsec}
\begin{figure}
     \includegraphics[scale=0.7]{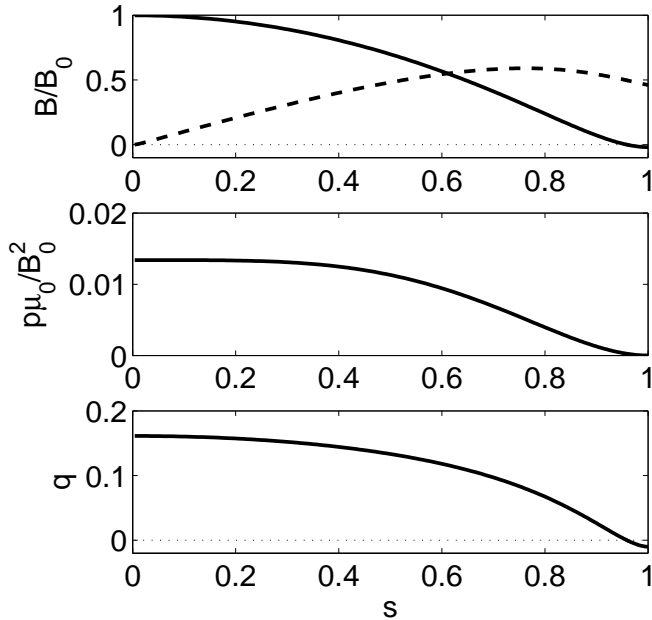} 
     \caption {{\it Equilibrium quantities as a functions of the normalized poloidal flux.  a) normalized equilibrium magnetic field components (solid -toroidal component, dashed - poloidal component), b) normalized pressure, c) safety factor }}
     \label{equil_profs}
    \end{figure}

     Equilibria closely modelling the RFP device RFX \cite{sonato_fid_03_1} is used. Circular plasma cross-section is taken with major radius $R_0=2.0$ m and aspect ratio $\varepsilon=0.23$. The RFP equilibrium parameters are $F\equiv\frac{B_\phi (a)}{<B_\phi>}$=-0.06, $\Theta\equiv\frac{B_\theta (a)}{<B_\phi>} $=1.41, where $B_\theta (a)$,$B_\phi (a)$ are the values of toroidal and poloidal components of the equilibrium magnetic field at the plasma edge and $<..>$ means averaging over the plasma column. The plasma current value is $I_p$=1.6 MA, on-axis toroidal field $B_0$=1.53 T. The poloidal beta value is $\beta_{pol}$=4$\%$, the on-axis values of electron and ion temperatures and densities are: $T_{e0}$=1 kEv, $T_{i0}$=400 eV, $n_{e0}=n_{i0}$=2.0$\cdot$ 10$^{19}$ m$^{-3}$. The pressure profile $p(r)$ is given by the expression $p(r)=n(r)*T(r)$, where $n(r)=n_0(1-(r/a)^6)$, $T(r)=T_0(1-(r/a)^3)$. The on-axis and edge safety factor values are: $q_0$=0.161, $q_a$=-0.01. In Fig. \ref{equil_profs} the poloidal and toroidal components of the magnetic field, pressure profile, and safety factor profile are shown for the described equilibrium as a functions of the normalized poloidal flux.
    
     \subsection{RWM stability without rotation}
\begin{figure}
     \includegraphics[scale=0.7]{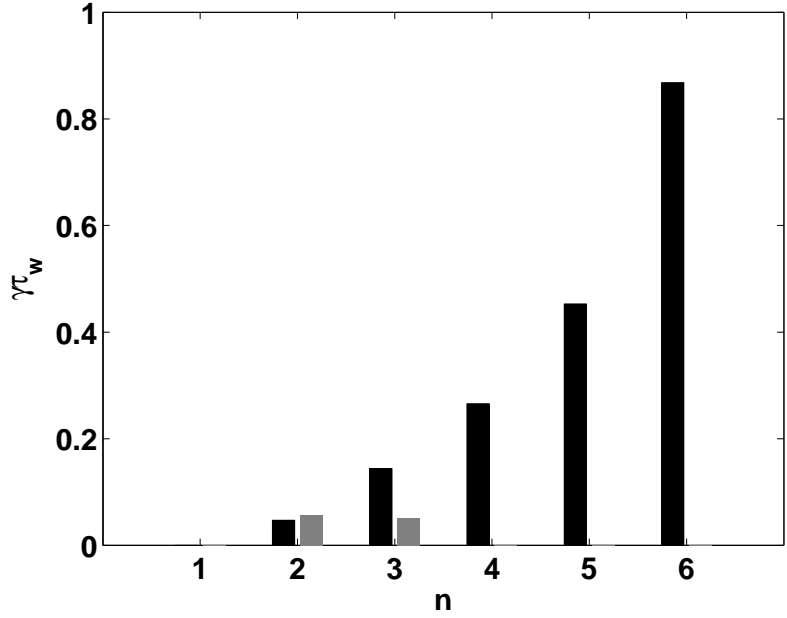} 
     \caption {{\it Spectrum of unstable RWMs. Black - $n<0$, grey - $n>0$.}}
     \label{n_spec}
    \end{figure}

\begin{figure}
     \includegraphics[scale=0.7]{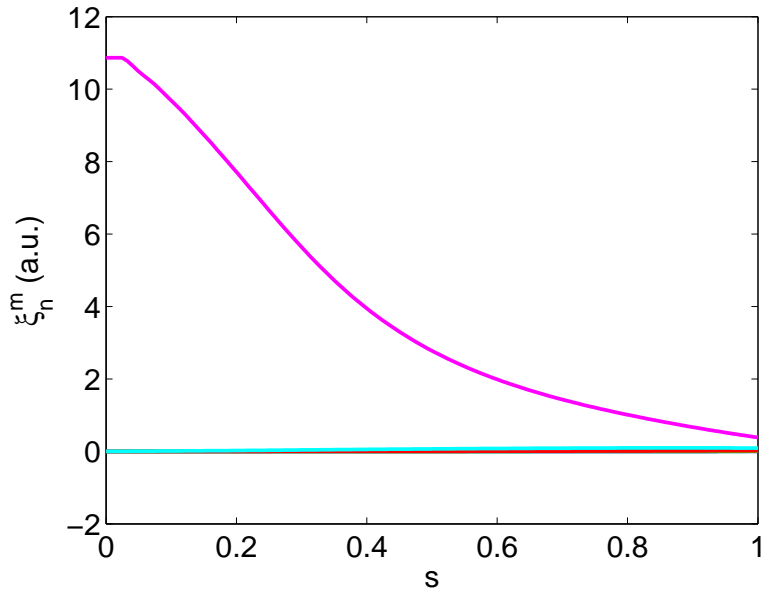} 
     \caption {{\it Poloidal harmonics ($m$=-5..5) of the RWM with n=-6. Dominant m=1 poloidal harmonic is shown in magenta solid line. All other harmonics are not visible.}}
     \label{n6_struc}
    \end{figure}

     The RWM spectrum in RFP is usually characterized by the presence of several unstable RWMs with different toroidal mode numbers $n$. Fourier harmonics with different helicities (positive and negative toroidal mode numbers) are visible in the spectrum due to the edge reversal of the equilibrium toroidal field. In Fig. \ref{n_spec} the spectrum of unstable RWMs is shown calculated for the equilibrium described in the previous section .
      The poloidal structure (different poloidal Fourier harmonics) of the normal displacement $\xi_n$ of the most unstable toroidal harmonic (n=-6) are shown in Fig. \ref{n6_struc}. It is observed that the harmonic with m=1 is dominant. This result points to the weak poloidal coupling in the RFP (differently from the tokamak configuration). In the further discussion the stability of the Fourier harmonic with n=-6 will be discussed if not stated otherwise.
     \subsection{Ideal MHD stability with plasma rotation. Continuum resonances} \label{cont_res}
\begin{figure}
     \includegraphics[scale=0.7]{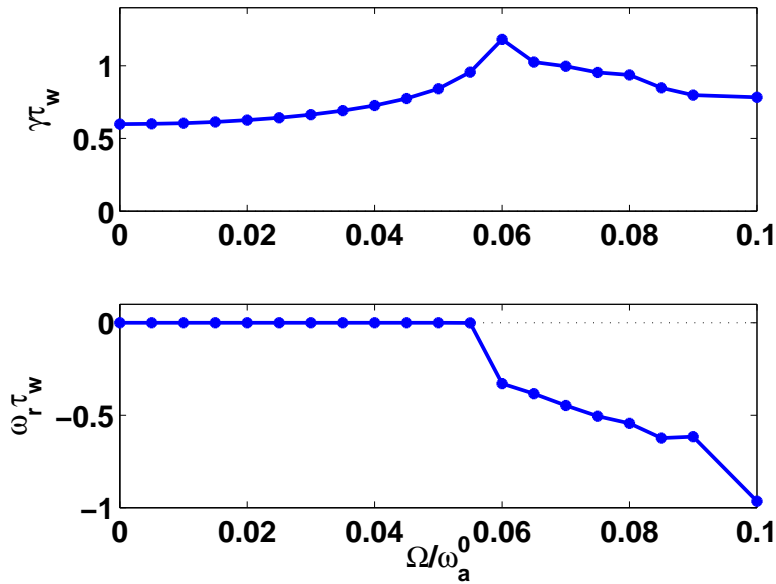} 
     \caption {{\it Mode resonance with the Alfv\'{e}n continuum. Complex eigenvalue (mode growth rate - $\gamma \tau_w$ and mode frequency $\omega_r \tau_w$) dependence on the plasma rotation $\Omega/\omega_a^^0$ normalized to the on-axis value of Alfv\'{e}n frequency.}}
     \label{al_res}
    \end{figure}

     Plasma rotation opens the possibility for the interaction of the static non-resonant RWM with the stable waves in the plasma. In ideal MHD such interaction is due to the mode resonance with continuum spectrum. Two continuum spectra are known to be resonant with RWM: Alfv\'{e}n continuum and sound continuum. The condition for the resonance appearance can be written in general as  $n\Omega+\omega_r=\omega_c$ where $\Omega$ is the plasma rotation frequency, $\omega_r$ is the RWM rotation frequency and $\omega_c$ is the continuum frequency. For the static RWM ($\omega_r=0$) this condition becomes $n\Omega= \omega_c$.  RWM resonance with the Alfv\'{e}n continuum spectrum appears when the condition $\vert n\Omega\vert =\vert \omega_{ca} \vert \equiv \vert k_\| v_a \vert$ is satisfied. Here $v_a\equiv B/(\mu_0\rho)^{1/2}$ is the Alfv\'{e}n velocity and $k_\|\equiv (m/q-n)$ is the parallel component of the RWM wave vector. In Fig. \ref{al_res} the mode growth rate $\gamma$ and mode rotation frequency $\omega_r$ (both normalized by the wall time $\tau_w$) dependence is shown on the plasma rotation frequency $\Omega$ (normalized by the Alfv\'{e}n frequency on the plasma axis $\omega_a^0$). The pressureless equilibrium with the equilibrium parameters $F,\Theta$ given in sec \ref{eqsec} is used.   
     
     Because of the non-resonance nature of the RWM in RFP, no resonance between the mode and the Alfv\'{e}n waves is possible at vanishing or slow plasma rotation.  The obtained  value of the plasma rotation frequency that starts to give the resonance with Alfv\'{e}n wave (seen at peak of the growth rate $\gamma\tau_w$) is $\Omega/\omega_a^0 \approx $0.06. This critical value agrees very well with the analytical prediction that is calculated using the resonance condition given above. The RWM growth rate $\gamma\tau_w$ starts to decrease for plasma rotation frequencies $\Omega>\omega_{ca}$ accompanied by the finite mode frequency appearance. This points to the damping effect appearance from the mode-continuum resonance, although no compete RWM stabilization is seen for the studied $\Omega$ range. The obtained value of the plasma rotation frequency for the mode resonance with the Alfv\'{e}n continuum is much less that the one obtained in the previous studies \cite{guo_phys_plasmas_99_1} due to the fact that the internal RWM with $n=-6$ is used in the present studies with the resonant surface closer to the plasma (and therefore smaller $k_\Vert$). It should be noted that the experimentally observed plasma rotation frequency in RFX is still much smaller than the value obtained here for the mode resonance with Alfv\'{e}n continuum ($\Omega$ value is of order of $10^{-3}\omega_a$ \cite{guazotto_ppcf_09_1}).
     
\begin{figure}
     \includegraphics[scale=0.7]{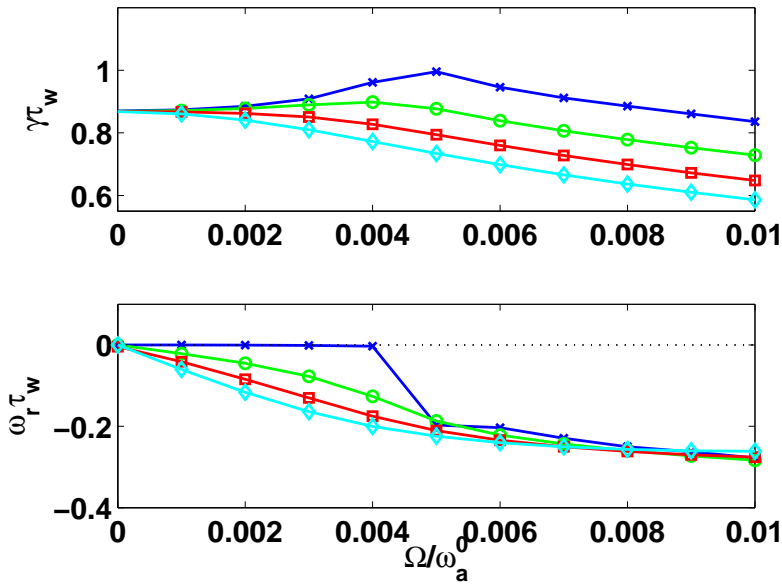} 
     \caption {{\it Complex eigenvalue (mode growth rate - $\gamma \tau_w$ and mode frequency $\omega_r \tau_w$) dependence on the plasma rotation $\Omega/\omega_a$ normalized to the Alfv\'{e}n frequency. Blue (and $\times$)- $\kappa$=0.0, green (and $o$) - $\kappa$=0.5, red (and $\square$) - $\kappa=1.0$, cyan (and $\lozenge$) - $\kappa=1.5$}}
     \label{so_res}
    \end{figure}

     The resonance of the static RWM with sound continuum appears when  $n\Omega=\omega_{cs}\equiv k_\|v_s$ where $v_s=(p/\rho)^{1/2}$ is the sound velocity. The behaviour of the complex eigenvalue in the plasma rotation frequency range corresponding to the resonance condition is shown in Fig. \ref{so_res} (line with $\times$) where an equilibrium with finite pressure is chosen. The obtained plasma frequency value that corresponds to the resonance  is $\Omega/\omega_a^0 \approx$ 0.004 and also agrees well with the value given by the analytical predictions. Note also that $\omega_{cs}\ll \omega_{ca}$. 
     
According to the studies performed in \cite{bondeson_ph_fluids_89_1} the resonant behaviour seen at $\omega_{cs}$ is not physical due to the fact that the ideal MHD model is unable to describe accurately the particle motion along the field lines. A reasonable damping model can be introduced \cite{chu_phys_plasmas_95_1} to remove the unphysical resonant behaviour observed in Fig. \ref{so_res} by damping the parallel sound wave in the ideal MHD model. This parallel wave is damped physically by ion Landau damping mechanisms. In this work viscous damping model is used (see Sec. \ref{sec2}) where a numerical coefficient $\kappa$ is introduced to measure the strength of the parallel sound wave damping. In Fig. \ref{so_res} the complex eigenvalue behaviour is shown for the different values of the sound wave damping coefficient.  
     It is seen that the the large $\kappa$ values (strong damping) removes the resonant behaviour of the RWM growth rate. At the same time viscous damping results in RWM suppression, with the larger suppression rate for the larger $\kappa$ values.

     \subsection{Kinetic resonances} \label{kin_res}
     A more physically consistent description of RWM stability for the low values of the plasma frequency is obtained considering the mode interaction with plasma particle drift motions.  Mode particle interaction is introduced in the MHD model through the pressure tensor \eqref{pres_tensor} and is included into the resonant operator \eqref{res_oper}:
   \begin{equation*}
     \lambda=\sum_{e,i}\sum_{m,l}\sum_\alpha\frac{n(\omega_{*N}+(\epsilon_k-3/2)\omega_{*T}+\Omega)-\omega}{n\omega_d+ n\Omega+[\alpha(m+nq)+l]\omega_b-i\nu_{eff} -\omega}
     \end{equation*}  
     where sum is over the poloidal Fourier harmonics $m$, bounce harmonics $l$, particle fraction $\alpha$, and particle species ($e,i$). 
      It is seen that resonance with several particle motions is possible depending on the value of the plasma rotation frequency. In these studies the mode resonance with precession drift and bounce motions is considered. 
      
\begin{figure}
     \includegraphics[scale=0.7]{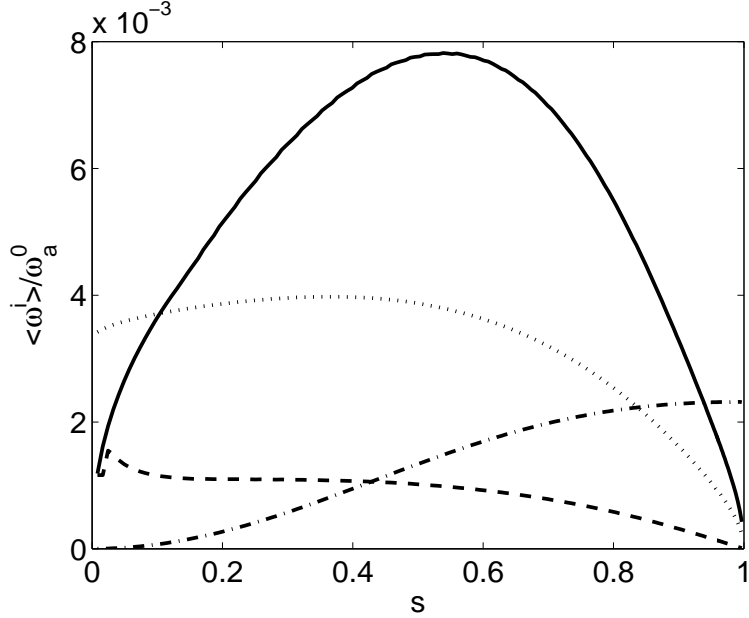} 
     \caption {{\it Ion kinetic frequencies averaged over the pitch angle $\Lambda$ and particle kinetic energy $\epsilon_k$. Solid line -$\omega_b$*0.1, dotted line - $\omega_p$*0.01, dashed line -$\omega_d$, dashed-dotted line - $\omega_*$. All frequencies are normalized to the central Alfv\'{e}n frequency value$\omega_a^0$.}}
     \label{dfreq}
    \end{figure}

    Full picture of the mode-particle resonance is complex due to the fact that contribution from the both trapped and passing fractions of the different plasma species (electrons and ions) should be taken into account. Moreover, particular particle motion (characterized by the motion's frequency value) have in general complex dependence on the pitch angle $\Lambda$ and particle kinetic energy  $\epsilon_k$ on each flux surface. Total effect of the RWM-particle resonances on the mode stability is obtained numerically considering mentioned dependencies. Kinetic frequencies averaged both over $\epsilon$ and $\Lambda$ are discussed in this section. Such simplified approach allows to make some qualitative predictions for the mode-particle resonances. The average kinetic frequencies for the ion plasma component are shown in Fig. \ref{dfreq} as functions of the normalized poloidal flux. 
    The frequency for the passing particles $\omega_p$ is equal to $\omega_b$ when $\alpha=1$ in \eqref{res_oper}. Note that $\omega_p$ and $\omega_b$ are scaled (by factors 0.01 and 0.1 respectively) to be able to compare profiles of different kinetic frequencies on one figure. The value of $\omega_*$ is the sum of the diamagnetic frequencies due to the density and temperature gradients. The plasma rotation frequency is not shown on the figure above, but comparing the experimental value ($\Omega\sim 10^{-3}\omega_a$ \cite{guazotto_ppcf_09_1}) with the calculated results  it is seen that $\Omega \sim\omega_d$.  The following approximate frequency ordering is obtained for the studied case:
    \begin{equation}
      \omega_r\ll\Omega\sim\omega_d\sim\omega_*\ll\omega_b \sim\omega_{ca}\sim\omega_s^{i0}<\omega_p<\omega_a^0
      \label{freq_order}
    \end{equation}
       where $\omega_s^{i0}\simeq 0.1 \omega_a^0$ is the central ion sound frequency. It is seen that the ion bounce frequency value is in the range both with the sound frequency and also with the frequency of the mode resonance with Alfv\'{e}n continuum. 
    
    In the present work the plasma frequency range $\Omega \lesssim \omega_s^{i0}$ is studied.  Two specific frequency subregions can be deduced following the frequency ordering obtained above: $\Omega\sim\omega_d$ and $\omega\sim\omega_b$. In each of this frequency region the plasma rotation frequency is close to the certain kinetic frequency (precession drift of bounce frequencies) and therefore the mode-particle resonance is expected to affect the RWM stability. It is useful for the further studies to have qualitative estimation of particle fractions and particle species that possibly could be involved in the mode-particle resonance in the mentioned frequency regions. The following qualitative analysis aims the estimation of the dominant terms of the resonant operator $\lambda$ for the two frequency subregions for different values of $\alpha$ (particle trapping) and $l$ (bounce harmonic number). 
 In the subregion   $\Omega \sim \omega_d$ the following conditions could lead to the mode-particle resonance: 
 
- $\alpha=0$. In this case $\lambda \simeq \frac{\omega_*+\Omega}{\omega_d+\Omega+(l/n)\omega_b}$. For $l=0$ mode resonance with precession drift motion has dominant effect. Both electrons and ions contribute. For $l \neq 0$ bounce motion could contribute to the mode-particle resonance for the slow rotation when  $\omega_d\sim(l/n)\omega_b$. As $\omega_d/\omega_b\approx 0.02$ across most of the plasma column (see Fig. \ref{dfreq}) condition $l\ll n$ should be satisfied that is not the case in the present studies ($min\vert (l/n) \vert=1/6$). Note that $\omega_d\sim\omega_b$ is seen locally in the plasma core or the plasma edge regions and therefore bounce motion can affect the RWM stability in these regions.
 
- $\alpha=1$. In this case resonant operator has general form \eqref {res_oper}. Passing particles can contribute to the mode-particle resonance for the slow plasma rotation when $(m+nq+l)/n  \sim \omega_d/\omega_p$. As $\vert \omega_d/\omega_p \vert<10^{-2}$ in the present studies the contribution from the passing particles can be expected when $m=-l, \vert q \vert \ll 1.0$ that is satisfied locally near the resonant surface. 

The main effect on the RWM stability in the low rotation frequency subregion is expected to come from the mode resonance with the precession drift frequencies. The effect from the bounce motion or passing particles is local.


Similar analysis for $\Omega \sim \omega_b$ gives:

- $\alpha=0$  As from the frequency ordering $\omega_d \ll \omega_b$ mode resonance only with bounce motion is possible, i.e. when $l \neq 0$. Then $\lambda \simeq \frac{\omega_*+\Omega}{(l/n)\omega_b+\Omega}$. The mode-particle resonance condition is satisfied when $(l/n)\omega_b\simeq -\Omega$. For present studies ($n=-6$) bounce harmonic with $l=6$ should have the dominant contribution. Note also that this condition could give the contribution to the RWM stability in the intermediate values of the plasma rotation frequency (i.e $\omega_d\lesssim\Omega<\omega_b$). This happens when $l/n<1$. Only ion component is involved as the electron bounce frequency is much higher.

- $\alpha=1$ In this case $\lambda\simeq \frac{\omega_*+\Omega}{[(m+nq+l)/n]\omega_p+\Omega}$. Passing particles can contribute to the mode resonance with the bounce motion when $(m+nq+l)/n  \sim \omega_b/\omega_p$. As  $\vert \omega_b/\omega_p \vert \sim 10^{-1}$ in this studies the contribution of the passing particles can be expected for $l=-m, \vert q \vert \lesssim 0.1$ that points to the possibility of the global effect from the passing particles (note that $\vert q \vert <0.16$ in the present studies). 


 Both trapped and passing particles could contribute to the mode-particle resonance for the plasma rotation in the order of ion sound frequency (note that $\omega_s\sim \omega_b$ from the frequency ordering \eqref{freq_order}). Trapped particle contribution comes from the bounce motion, passing particles effect could be global (note that this is different from the case of the low rotation frequencies $\Omega\sim\omega_d$).

     \subsection{Comparison of the viscous and kinetic damping models}
\begin{figure}
     \includegraphics[scale=0.7]{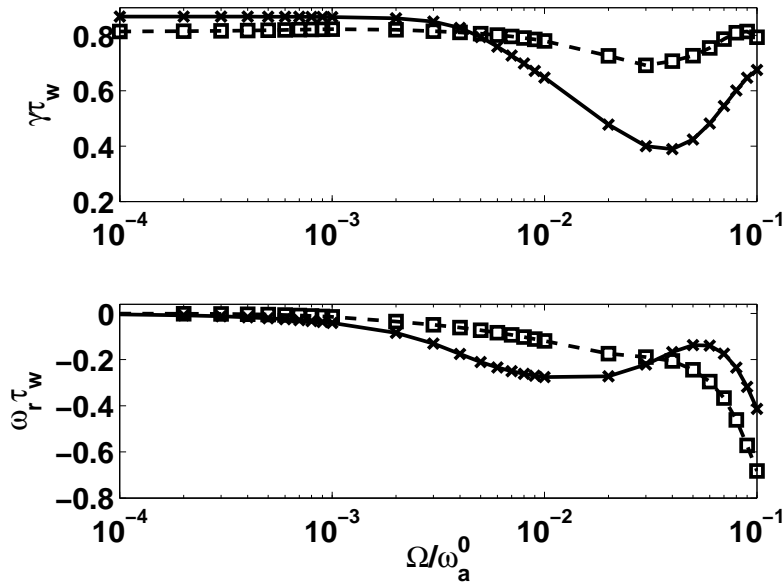} 
     \caption {{\it Complex eigenvalue (mode growth rate - $\gamma \tau_w$ and mode frequency $\omega_r \tau_w$) dependence on the plasma rotation $\Omega/\omega_a$ normalized to the Alfv\'{e}n frequency. Solid line with $\times$- fluid damping (sound wave damping with $\kappa=1.0$), dashed line with $\square$ - model with kinetic resonances included.}}
     \label{comp_fl_kin}
    \end{figure}

     It was mentioned above that the reasonable damping model can be used in ideal MHD description to model physical ion Landau damping. Viscous damping model is used here with the damping coefficient $\kappa$=1.0 (strong damping) for the comparison with the MHD description including kinetic effects. The results of comparison are shown in Fig. \ref{comp_fl_kin}. It is seen that kinetic damping model gives slightly lower growth rate for the low plasma rotation frequencies ($\Omega\lesssim \omega_d$) while the higher suppression rate is obtained for $\Omega \sim \omega_b$ using the viscous damping model. Increase of the RWM growth rate for $\Omega>5\cdot 10^{-2}\omega_a^0$ is explained by the resonant RWM behaviour in vicinity of the mode resonance with Alfv\'{e}n continuum (see frequency ordering \eqref{freq_order}). 
     \subsection{Different kinetic resonances}
\begin{figure}
     \includegraphics[scale=0.7]{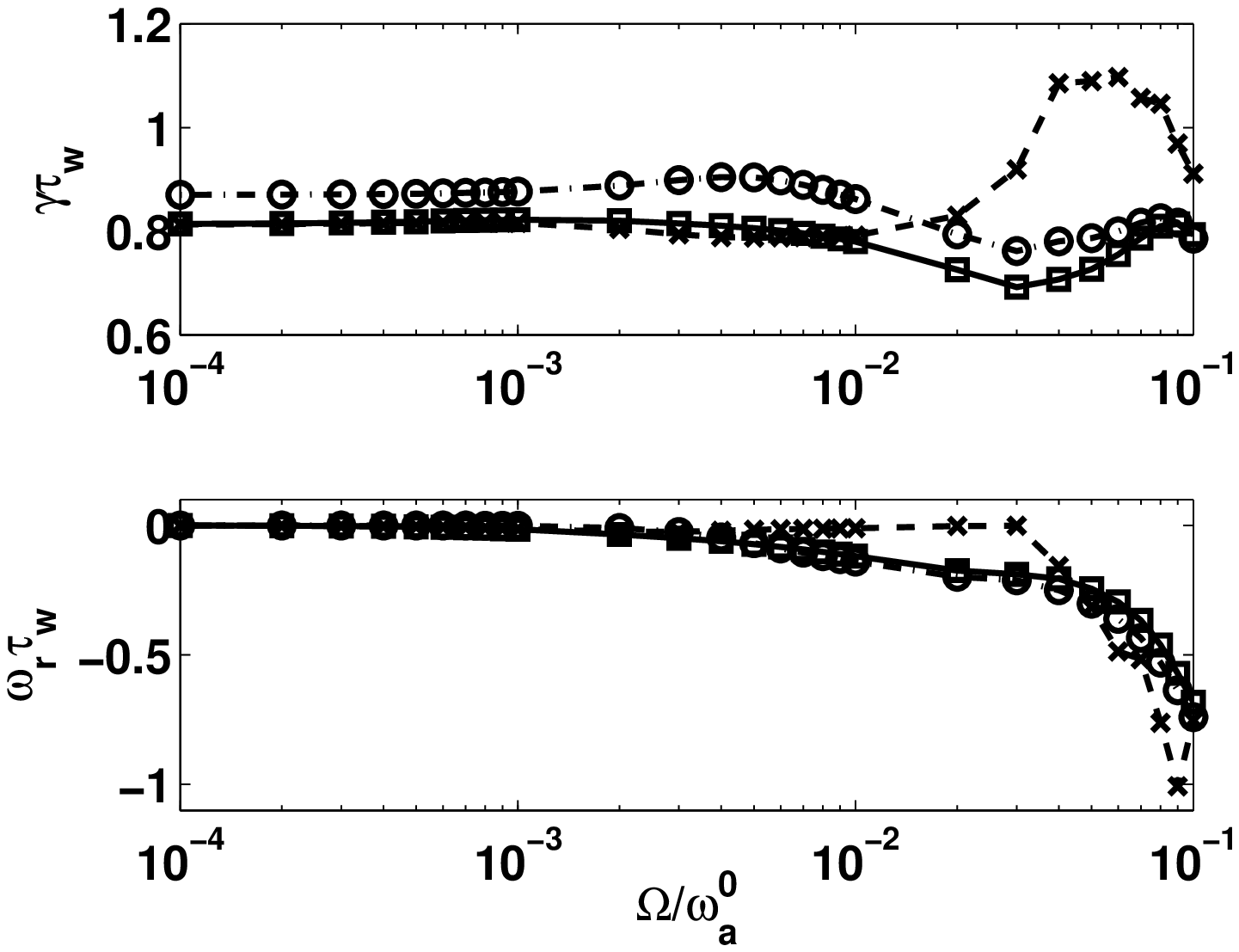} 
     \caption {{\it Complex eigenvalue (mode growth rate - $\gamma \tau_w$ and mode frequency $\omega_r \tau_w$) dependence on the plasma rotation $\Omega/\omega_a$ normalized to the Alfv\'{e}n frequency. Solid line with $\square$ -full model, dashed line with $\times$ - resonance only with $\omega_d$ is considered, dashed-dotted line with $o$ - resonance only with $\omega_b$ is considered. Both trapped and passing particles are included}}
     \label{kin_res_comp1}
    \end{figure}

     Several kinetic frequencies are included in the resonant operator \eqref{res_oper} and therefore can contribute to the total effect on the RWM stability. In the previous section 'full' model is used taking into account mode resonance with precession drift and bounce motions for trapped and passing particles. In order to evaluate the importance of the particular kinetic resonance and particle fraction on the total effect the following studies are performed. It can be seen from the definition of the resonant operator that the frequencies of the kinetic resonances are included in additive manner, giving the possibility to exclude all but one resonance from the calculations. Particle fractions (trapped or passing) can be also treated separately. In Fig. \ref{kin_res_comp1} the 'full' model (blue curve) is compared to the 'only bounce' model (red curve) when only bounce frequencies are taken into account  and with 'only precession drift' model (green curve) when only precession drift frequencies are taken into account. Both trapped and passing particles are included.
    Two plasma rotation frequency regions can be distinguished separated by the plasma rotation frequency value $\Omega/\omega_a^0 \approx 10^{-2}$. For the region $\Omega/\omega_a^0 < 10^{-2}$ eigenvalue calculated for the 'full' model case is closely follows that calculated for the 'only precession drift' model. Note that the mode growth rate for these two cases is slightly less than that for the 'only bounce' model. For the region $\Omega/\omega_a^0 > 10^{-2}$ situation becomes opposite i.e. the eigenvalue trace corresponding to the 'full' model follows that for the 'only bounce' model.  Resonant behaviour of the RWM growth rate is seen for the 'only precession drift' model in this frequency range. It is attributed to the mode resonance with the Alfv\'{e}n continuum spectrum.
 
\begin{figure}
     \includegraphics[scale=0.7]{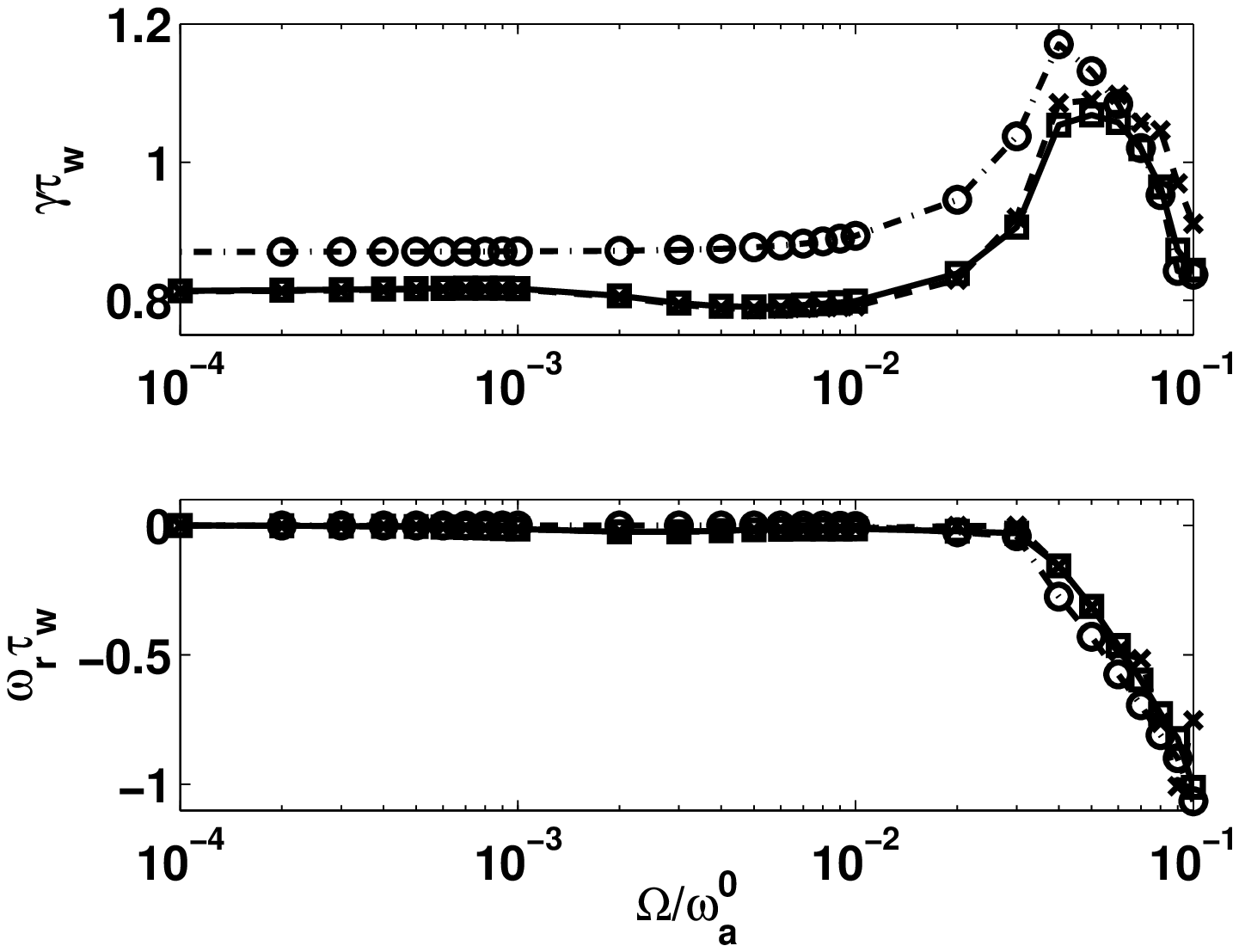} 
     \caption {{\it Complex eigenvalue (mode growth rate - $\gamma \tau_w$ and mode frequency $\omega_r \tau_w$) dependence on the plasma rotation $\Omega/\omega_a$ normalized to the Alfv\'{e}n frequency. Solid line with $\square$ -full model, dashed line with $\times$ - resonance only with $\omega_d$ is considered, dashed-dotted line with $o$ - resonance only with $\omega_b$ is considered. Only trapped particles are included}}
     \label{kin_comp2}
    \end{figure}

    In Fig. \ref{kin_comp2} the same resonances are compared where only trapped particles are included. It is seen that the mode growth rate behaviour is different in the second plasma rotation frequency region $\Omega/\omega_a^0 > 10^{-2}$ for the 'full' and 'only bounce' models. Now similar resonant behaviour is seen for all three models at $\Omega/\omega_a^0 \approx 0.05$. The growth rate for the 'full' model is almost identical with that for the 'only precession drift' model for the whole range of the studied plasma rotation frequencies. 
   The effect of the mode resonance with different particle fractions for two frequency regions ($\Omega/\omega_a^0 < 10^{-2}$ and  $\Omega/\omega_a^0 > 10^{-2}$ ) can be seen. In the region $\Omega/\omega_a^0 < 10^{-2}$ the effect of the trapped and passing particles on the RWM stability is observed.  RWM is suppressed due to the resonance with the precession drift motion ($\times$ curve in Fig. \ref{kin_res_comp1}) and destabilized due to the resonance with the passing particles($o$ curve in Fig. \ref{kin_res_comp1}). This two effects almost cancel each other as it is seen from the behaviour of the 'full' model eigenvalue ($\square$ curve in Fig. \ref{kin_res_comp1}). Note that there is no effect from the bounce motion in this frequency region (compare $o$ curves in Figs.\ref{kin_res_comp1} and \ref{kin_comp2}). In the region $\Omega/\omega_a^0 > 10^{-2}$ only passing particles have the effect on the RWM stability. Indeed, similar eigenvalue behaviour for the three studied cases seen in Fig. \ref{kin_comp2} points (somewhat surprisingly) that the mode resonance with bounce motion does not affects the RWM stability. The resonant behaviour and subsequent mode suppression seen for  $\Omega/\omega_a^0 > 5\cdot10^{-2}$ is attributed to the mode resonance with the Alfv\'{e}n continuum. On the other hand substantial mode suppression is observed in Fig. \ref{kin_res_comp1} ($o$ and $\square$ curves) in this frequency range that effectively prevents the mode growth due to the resonance with Alfv\'{e}n continuum. Note that although the frequency of the passing particles is in general much higher than the frequencies of the bounce and precession drift motions, the possibility of the mode resonance with the passing particles for low frequencies exists and the conditions for such interaction are qualitatively estimated in Sec. \ref{kin_res} (cases $\alpha=1$ for bounce and precession drift frequency ranges).
      
      \subsection{Equilibrium plasma pressure}
      In the previous studies of the kinetic effects for tokamak configuration \cite{liu_phys_plasmas_08_2} it was shown that equilibrium pressure affects the mode interaction with the particle motions. Both plasma pressure profile and absolute value can change the behaviour of the complex eigenvalue. In order to investigate the effect of pressure on the mode resonance with the plasma particles for RFP configuration the studies of the plasma pressure profile pressure absolute value (characterized by the $\beta_p$ parameter) are performed below.
        \subsubsection{Pressure profile}
\begin{center} 
        \begin{figure}
     \includegraphics[scale=0.7]{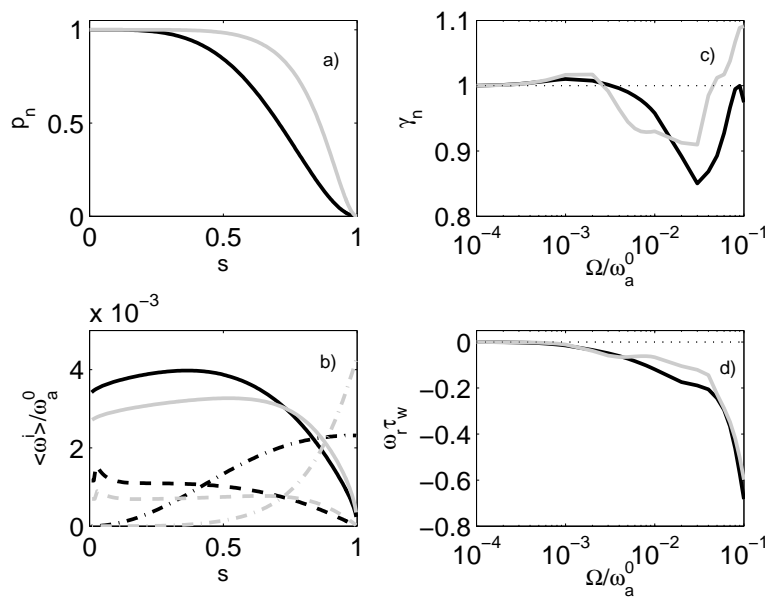} 
     \caption {{\it Results of the pressure profile studies. a) normalized pressure radial profiles; b) kinetic frequencies (solid - $\omega_p*0.01$, dashed - $\omega_d$, dashed-dotted - $\omega_*$), $\omega_b$ profiles are not shown as no effect on RWM stability is found from the mode resonance with the bounce motion ; c) RMW growth rate dependence on the plasma rotation frequency; d) RWM rotation frequency dependence on the plasma rotation frequency. Black - $n(r)=n_0(1-(r/a)^6,T(r)=T_0(1-(r/a)^3$; grey - $n(r)=n_0(1-(r/a)^{12},T(r)=T_0(1-(r/a)^6$;}}
     \label{dif_pres}
    \end{figure} 
    \end{center}

         Pressure profile shape affects the value of the pressure derivative and therefore changes the values of $\omega_*$ frequencies and also (according to \cite{liu_phys_plasmas_08_2}) affects the precession drift frequency term of the resonance operator \eqref{res_oper}. Two pressure profiles are compared in the present studies: the one used for the studies above (see Sec.\ref{eqsec}) and more flat one that is characterized by the following density and temperature profiles: $n^{i,e}(r/a)=n_0^{i,e}(1-(r/a)^12)$,$T^{i,e}(r/a)=T_0^{i,e}(1-(r/a)^6)$. The results of comparison are shown in Fig. \ref{dif_pres}. It is see that by changing pressure profile the diamagnetic frequency profile is changed (dashed-dotted line in Fig. \ref{dif_pres}b). Note also the slight change of $\omega_d$ (dashed line) and $\omega_p$(solid line) profiles. The RWM is suppressed more (Fig. \ref{dif_pres}c) in the region $\Omega\sim\omega_d$ for the case of the flat pressure profile, but for $\Omega\sim\omega_b$ more peaked profile gives better suppression. RWM destabilization is seen for $\Omega\gtrsim 5\cdot 10^{-2}\omega_a^0$ in both cases attributed to the effect from the mode resonance with the Alfv\'{e}n continuum.
    
    \subsubsection{Effect of $\beta_p$}
\begin{center} 
        \begin{figure}
     \includegraphics[scale=0.7]{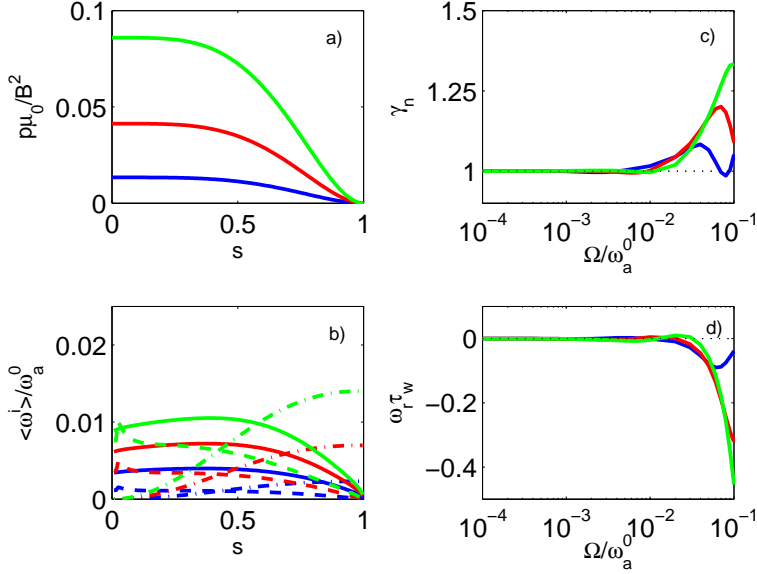} 
     \caption {(Colour online) {\it Results of the pressure value studies for the Fourier component with $n=-5$. a) normalized pressure radial profiles; b) kinetic frequencies (solid - $\omega_p*0.01$, dashed - $\omega_d$, dashed-dotted - $\omega_*$); c) RMW growth rate dependence on the plasma rotation frequency; d) RWM rotation frequency dependence on the plasma rotation frequency. Blue - $\beta_p$=5\%; red - $\beta_p$=15\%, green -$\beta_p$=25\%;}}
     \label{kin_pr_val}
    \end{figure} 
    \end{center}

      The absolute value of the plasma pressure (characterized usually by the beta value) is one of the most important parameters in the fusion research. Operation with the high beta value is necessary for the efficient operation of the fusion reactor. In this work poloidal beta $\beta_p\equiv\frac{8\pi<p>}{I_p^2}V_{tot}$ is used for the pressure value characterization, where $<p>$ is the equilibrium plasma pressure averaged over the plasma volume $V_{tot}$, $I_p$ - plasma current. 
 It is seen from the frequency ordering \eqref{freq_order} that for $n=-6$ Fourier harmonic the effects from the kinetic and continuum resonances occurs for the same plasma rotation value $\omega\sim\omega_s^{i0}$.  In order to separate these effects another unstable Fourier harmonic is chosen with $n$=-5 for which the mode resonance with the Alfv\'{e}n continuum is shifted towards higher plasma rotation frequencies. Indeed using analytical expression (see Sec. \ref{cont_res}) the mode resonance with Alfv\'{e}n continuum appears for this harmonic, when $\Omega\approx 0.2\omega_a^0$. The results of the studies for three $\beta_p$ values ($\beta_p$=5\%,15\%,25\%) are shown in Fig. \ref{kin_pr_val}
    It is observed (Fig. \ref{kin_pr_val}a) that the values of kinetic frequencies depend on poloidal beta. Dependence for different kinetic frequencies can be approximated as: $\omega_p(\beta_p)\sim \sqrt{\beta_p}$, $\omega_d(\beta_p)\sim \beta_p$, $\omega_*(\beta_p) \sim \beta_p$ that is similar as for the large aspect ration scaling for tokamak equilibrium \cite{liu_phys_plasmas_08_2}. Change of the kinetic frequencies magnitude causes also the change of the plasma rotation frequency of the mode-particle resonances. The consequence of such resonance shift is that the stabilizing effect from the kinetic resonances in the range $\Omega\sim\omega_b$ is seen only for the low value of poloidal beta (blue and red curves In Fig. \ref{kin_pr_val}c) for the studied range of plasma rotation frequencies. The mode destabilization seen for the $\Omega/\omega_a^0\gtrsim 0.01$ is similar for all cases and can be attributed to the effect of the mode resonance with Alfv\'{e}n continuum. Note that no such substantial shift of the kinetic frequencies is seen for tokamaks, due to the fact that the beta range where unstable RWM is observed is of the order of one percent that is much smaller than in RFP. 
      
        \subsection{Mode resonance with electron and ion plasma components for $\Omega\sim\omega_d$}
\begin{figure}
           \includegraphics[scale=0.7]{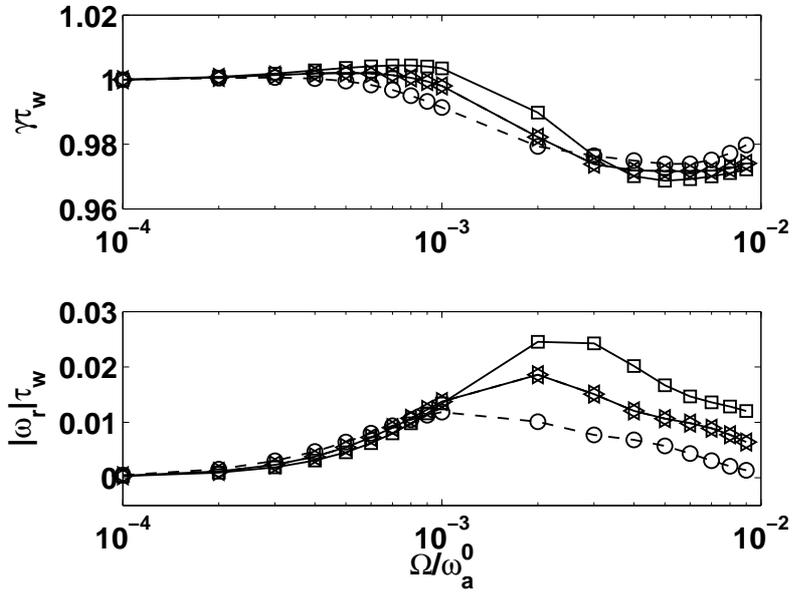}
           \caption {{\it Complex eigenvalue dependence on the plasma rotation frequency for the Fourier component with $n=-6$. Solid line with $\square$ - $\Omega>0,C_p=0.33$, solid line with $\lhd$ - $\Omega>0,C_p=0.5$, dashed line with $\diamondsuit$ - $\Omega<0, C_p=0.33$, dashed line with $\rhd$ - $\Omega<0,C_p=0.5$. Mode growth rate $\gamma$ for each curve is normalized to the value for $\Omega=10^{-4}$.}}     
          \label{dif_rot}
        \end{figure}

        It was shown in \cite{liu_phys_plasmas_08_1} for tokamak that temperature difference between electron and ion components can sufficiently affect the RWM stability for the low plasma rotation frequencies ($\Omega\sim\omega_d$) pointing to the different contribution from the mode resonance with electrons and ions. In general, contribution both from ion and electron components to the total effect is present at each flux surface due to the specific dependence of the precession drift frequency on the pitch angle for each flux surface. In particular, $\omega_d$ changes sign as a function of $\Lambda$ (see Fig. 1 in \cite{liu_phys_plasmas_08_1}) and therefore mode resonance both with electrons and ions is possible on the same flux surface as $\omega_d$ depends on the sign of the particle charge (note that such $\omega_d(\Lambda)$ dependence makes possible RWM suppression with zero plasma rotation). However for the present studies it is seen (Fig. \ref{dfreq}) that the average ion precession drift frequency is positive for all radial points. This means that precession drift frequency for particular plasma component has predominantly one sign as a function of $\Lambda$ on each flux surface and therefore the contributions from electrons and ions could be separated by the sign of the plasma rotation (i.e. mode resonance only with ion or electron component will determine the total effect on the RWM stability for the particular plasma rotation direction). In Fig. \ref{dif_rot} the complex eigenvalue behaviour is shown as a function of the plasma rotation frequency in the low plasma rotation region ($\Omega\sim\omega_d$). The complex eigenvalue is calculated for the Fourier harmonic $n=-6$ taking into account only the mode resonance with the precession drift motion. Two values of the ion- electron temperatures ratio $C_p=\frac{T_0^i}{T_0^i+T_0^i}$ ( $C_p=0.33$ and $C_p=0.5$) are compared for the positive and negative plasma rotation directions. First $C_p$ value corresponds to the ion-electron temperature ratio used in the present work ($T_{e0}$=1 $keV$, $T_{i0}$=0.5 $keV$), second $C_p$ value corresponds to the case of equal central temperatures of electrons and ions. In this case $\omega_*^i=-\omega_*^e, \omega_d^i=-\omega_d^e$.       
       Absolute value of the mode rotation frequency ($\vert \omega_r \vert \tau_w$) is plotted in order to make comparison easier (mode frequency rotation changes sign when the plasma rotation direction is reversed). It is seen that the mode growth rate behaves differently for the positive and negative plasma rotation directions in the case when $C_p=0.33$. For negative $\Omega$ (dashed line with $\diamondsuit$) mode stabilization effect is observed for the lower $\Omega$ values than for the positive rotation direction (solid line with $\square$). This could mean that $\vert \omega_d^e \vert < \vert \omega_d^i \vert$ and the mode resonance with electron component occurs for the lower $\Omega$ values.  In the case when $C_p=0.5$ the same effect is observed for the positive and negative $\Omega$ directions (solid line with $\lhd$ and dashed line with $\rhd$ are overlapped). This is expected result as the kinetic frequency values are equal for $C_p=0.5$ and therefore the effect from the different plasma components is independent on the plasma rotation direction.

      \section{Discussion and conclusions}
      The results of the present investigations show that the kinetic effects do not change the RWM stability substantially for the studied RFP equilibrium. An attempt to explain such result could be made by considering the dispersion relation for the RWM written in terms of the fluid and kinetic energy components (see eq. 8 in \cite{hu_prl_04_1}). This dispersion relation is used for the studies of the kinetic effects via the perturbative approach. Qualitatively, the stability of RWM is determined by the relation between fluid and kinetic energy components. It was shown \cite{liu_phys_plasmas_09_1} that for the pressure driven RWM seen in tokamak configuration kinetic and fluid  terms are comparable. Therefore changes in kinetic term can substantially affect the overall stability of RWM. The calculations made for the studied here RFP equilibrium show that for the current driven RWM fluid term is much larger that the kinetic term. Therefore qualitatively kinetic term does not play an important role for the RWM stability in RFP. More work should be done in order to make the general conclusion about the role of the kinetic energy on the RWM stability in RFP. 
      
      It is also seen during the present studies, that the observed effect on the RWM stability from the kinetic resonances comes mainly from the mode resonance with the {\it passing particles} for the plasma rotation of the order of ion sound frequency. Surprisingly no effect from the trapped particles (bounce motion) is seen in this region. It is observed that also for the low rotation (of the order of precession frequency) mode resonance with the passing particles have quantitatively similar effect on RWM stability as compared to the mode resonance with the precession drift motion. 
    
           
      In conclusion the studies of the RWM interaction with particle motions was investigated for RFP configuration. The equilibrium close to the experimental was used. The effect from the kinetic resonances is small, but the eigenvalue behaviour for the low plasma rotation frequency is explained better by the model with kinetic resonances than the ideal MHD model. Two regions of stabilization was found similarly to the previous work[] that correspond to the different plasma particles.  

\begin{thebibliography}{1}
  \bibitem {garofallo_prl_99_1} A. M. Garofalo, A. D. Turnbull, M. E. Austin, J. Bialek, M. S. Chu, K. J. Comer, E. D. Fredrickson,R. J. Groebner, R. J. La Haye, L. L. Lao, E. A. Lazarus, G. A. Navratil, T. H. Osborne, B. W. Rice, S. A. Sabbagh, J. T. Scoville, E. J. Strait, and T. S. Taylor, Phys. Rev. Lett., {\bf 82}, 3811 (1999).
   \bibitem {gryaznev_baps_03_1} M Gryaznevich, C G Gimblett, T C Hender, D F Howell, S Pinches, Y Liu, A Bondeson, Bull. Am. Phys. Soc., {\bf 48}, 307 (2003).
   \bibitem{shilov_pop_04_1} M. Shilov, C. Cates, R. James, A. Klein, O. Katsuro-Hopkins, Y. Liu, M. E. Mauel, D. A. Maurer, G. A. Navratil, T. S. Pedersen, and N. Stillits, R. Fitzpatrick, S. F. Paul, Phys. Plasmas, {\bf 11}, 2573 (2004).
   \bibitem {alper_ppcf_89_1} B Alper, M K Bevir, H A B Bodin, C A Bunting, P G Carolan, J Cunnane, D E Evans, C G Gimblett, R J Hayden, T C Hender, A Lazaros, R W Moses, A A Newton, P G Noonan, R Paccagnella, A Patel, H Y W Tsui and P D Wilcock, Plasma Phys. Controlled Fusion, {\bf 31}, 205 (1989).
   \bibitem {brunsell_pop_03_1} P. R. Brunsell, J.-A. Malmberg, D. Yadikin, and M. Cecconello, Phys. Plasmas, {\bf 10}, 3823 (2003).
   \bibitem{bolzonella_32eps_1} T. Bolzonella, E. Martines, D. Terranova, P. Zanca, R. Cavazzana, L. Grando, N. Pomaro, G. Serianni, N. Vianello, M. Zuin, Proceedings of the 32nd EPS Plasma Physics Conference (Tarragona) ECA (European Physical Society, Mulhouse Cedex, France, 2005), Vol. 29C, p. 1.107.
   \bibitem {bishop_ppcf_89_1} C. M. Bishop, Plasma Phys. Controlled Fusion, {\bf 31}, 1179 (1989).
   \bibitem {fitzpatrick_pop_96_1} R. Fitzpatrick and T. H. Jensen, Phys. Plasmas {\bf 3}, 2641 (1996).
   \bibitem {okabayashi_nf_98_1} M. Okabayashi, N. Pomphrey and R.E. Hatcher, Nucl. Fusion, {\bf 38}, 1607 (1998).
   \bibitem {fitzpatrick_pop_99_1} R. Fitzpatrick and E. P. Yu, Phys. Plasmas, {\bf 6}, 3536 (1999).
   \bibitem {liu_pop_00_1} Y. Q. Liu, A. Bondeson, C. M. Fransson, B. Lennartson, and C. Breitholtz, Phys. Plasmas, {\bf 7}, 3681 (2000).
   \bibitem {pustovitov_ppr_01_1} V. D. Pustovitov, Plasma Phys. Rep.,  {\bf 27} 195 (2001).
   \bibitem{okabayashi_pop_01_1} M. Okabayashi, J. Bialek, M. S. Chance, M. S. Chu, E. D. Fredrickson, A. M. Garofalo, M. Gryaznevich, R. E. Hatcher, T. H. Jensen, L. C. Johnson, R. J. La Haye, E. A. Lazarus, M. A. Makowski, J. Manickam, G. A. Navratil, J. T. Scoville, E. J. Strait, A. D. Turnbull, and M. L. Walker, Phys. Plasmas, {\bf 8}, 2071 (2001).
   \bibitem{strait_pop_04_1} E. J. Strait, J. M. Bialek, I. N. Bogatu, M. S. Chance, M. S. Chu, D. H. Edgell, A. M. Garofalo, G. L. Jackson, R. J. Jayakumar, T. H. Jensen, O. Katsuro-Hopkins, J. S. Kim, R. J. La Haye, L. L. Lao, M. A. Makowski, G. A. Navratil, M. Okabayashi, H. Reimerdes, J. T. Scoville, A. D. Turnbull, and DIII-D Team, Phys. Plasmas, {\bf 11} 2505 (2004).
   \bibitem{sabbagh_prl_04_1} S. A. Sabbagh, R. E. Bell, J. E. Menard, D. A. Gates, A. C. Sontag, J. M. Bialek, B. P. LeBlanc, F. M. Levinton, K. Tritz, and H. Yuh, Phys. Rev. Lett., {\bf 97}, 045004 (2006).
   \bibitem{brunsell_prl_04_1} P. R. Brunsell, D. Yadikin, D. Gregoratto, R. Paccagnella, T. Bolzonella, M. Cavinato, M. Cecconello, J. R. Drake, A. Luchetta, G. Manduchi, G. Marchiori, L. Marrelli, P. Martin, A. Masiello, F. Milani, S. Ortolani, G. Spizzo, and P. Zanca,  Phys. Rev. Lett., {\bf 93}, 225001 (2004).
   \bibitem {pacaggnella_prl_06_1} R. Paccagnella, S. Ortolani, P. Zanca, A. Alfier, T. Bolzonella, L. Marrelli, M. E. Puiatti, G. Serianni, D. Terranova, M. Valisa, M. Agostini, L. Apolloni, F. Auriemma, F. Bonomo, A. Canton, L. Carraro, R. Cavazzana, M. Cavinato, P. Franz, E. Gazza, L. Grando, P. Innocente, R. Lorenzini, A. Luchetta, G. Manduchi, G. Marchiori, S. Martini, R. Pasqualotto, P. Piovesan, N. Pomaro, P. Scarin, G. Spizzo, M. Spolaore, C. Taliercio, N. Vianello, B. Zaniol, L. Zanotto, and M. Zuin, Phys. Rev. Lett., {\bf 97}, 075001 (2006).
  \bibitem{reimerdes_phys_plasams_06_1} H. Reimerdes, T. C. Hender, S. A. Sabbagh, J. M. Bialek, M. S. Chu, A. M. Garofalo, M. P. Gryaznevich, D. F. Howell, G. L. Jackson, R. J. La Haye, Y. Q. Liu, J. E. Menard, G. A. Navratil, M. Okabayashi, S. D. Pinches, A. C. Sontag, E. J. Strait, W. Zhu,1 M. Bigi, M. de Baar, P. de Vries, D. A. Gates, P. Gohil,R. J. Groebner, D. Mueller, R. Raman, J. T. Scoville, W. M. Solomon, the DIII-D Team, JET-EFDA Contributors, and the NSTX Team, Phys. Plasmas {\bf 13}, 056107 (2006).

  \bibitem{gregoratto_ppcf_01_1} D. Gregoratto, A. Bondeson, M. S. Chu and A. M. Garofalo, Plasma Phys. Controlled Fusion {\bf 43}, 1425 (2001).
  \bibitem{zheng_prl_05_1} L. J. Zheng, M. Kotschenreuther, and M. Chu, Phys. Rev. Lett. {\bf 95}, 255003 (2005).
  \bibitem{bondeson_prl_94_1} A. Bondeson and D. J. Ward, Phys. Rev. Lett. {\bf 72}, 2709 (1994).
  \bibitem{betti_prl_95_1} R. Betti, J. P. Freidberg, Phys. Rev. Lett. {\bf 74}, 2949 (1995).
  \bibitem{bondeson_ph_fluids_89_1} A. Bondeson and R. Iacono, Phys. Fluids B {\bf 1}, 1431 (1989).
  \bibitem{chu_phys_plasmas_95_1} M. S. Chu, J. M. Greene, T. H. Jensen, R. L. Miller, A. Bondeson, R. W. Johnson, and M. E. Mauel, Phys. Plasmas {\bf 2}, 2236 (1995).
  \bibitem{bondeson_phys_plasmas_96_1} A. Bondeson and  M. S. Chu, Phys. Plasmas {\bf 3}, 3013 (1996).
  \bibitem{reimerdes_prl_07_1} H. Reimerdes, A. M. Garofalo, G. L. Jackson, M. Okabayashi, E. J. Strait, M. S. Chu, Y. In, R. J. La Haye, M. J. Lanctot, Y. Q. Liu, G. A. Navratil, W. M. Solomon, H. Takahashi, and R. J. Groebner, Phys. Rev. Lett.  {\bf 98}, 055001 (2007).
  \bibitem{hu_prl_04_1} Bo Hu and R. Betti, Phys. Rev. Lett. {\bf 93}, 105002 (2004).
  \bibitem{polevoi_02_1} A.R. Polevoi, S.Yu. Medvedev, V.D. Pustovitov, V.S. Mukhovatov, M. Shimada, A.A. Ivanov, Yu.Yu. Poshekhonov, M.S. Chu Proceedings of the 19th Int. Conf. on Fusion Energy (Lyon) (IAEA, Vienna 2002), CD-ROM file CT/P-08
  \bibitem{jiang_phys_plasmas_95_1} Z. X. Jiang, A. Bondeson and R. Paccagnella, Phys. Plasmas {\bf 2}, 442 (1995).
  \bibitem{guo_phys_plasmas_99_1} S. C. Guo, J. P. Freidberg, and R. Nachtrieb, Phys. Plasmas {\bf 6}, 3868 (1999).
  \bibitem{sonato_fid_03_1} P. Sonato, G. Chitarin, P. Zaccaria, F. Gnesotto, S. Ortolani, A. Buffa, M. Bagatin, W. R. Baker, S. Dal Bello, P. Fiorentin, L. Grando, G. Marchiori, D. Marcuzzi, A. Masiello, S. Peruzzo, N. Pomaro and G. Serianni, Fusion Eng. Des. {\bf 66}, 161 (2003).
  \bibitem{liu_phys_plasmas_08_1} Y. Q. Liu, M. S. Chu, I. T. Chapman, and T. C. Hender, Phys. Plasmas {\bf 15}, 112503 (2008)
  \bibitem{lutjens_cpc_96_1} H. L\"{u}tjens, A. Bondeson, O. Sauter, Comp. Phys. Commun. {\bf 97}, 219 (1996).
  \bibitem{guazotto_ppcf_09_1} L. Guazzotto and R. Paccagnella, Plasma Phys. Controlled Fusion {\bf 51}, 065013 (2009).
  \bibitem{liu_phys_plasmas_08_2} Y. Q. Liu, M. S. Chu,C. G. Gimblett, and R. J. Hastie, Phys. Plasmas {\bf 15}, 092505 (2008).
  \bibitem{liu_phys_plasmas_09_1} Y. Q. Liu, I. T. Chapman, M. S. Chu, H. Reimerdes, F. Villone, R. Albanese, G. Ambrosino, A. M. Garofalo, C. G. Gimblett, R. J. Hastie, T. C. Hender,
G. L. Jackson, R. J. La Haye, M. Okabayashi, A. Pironti, A. Portone, G. Rubinacci, and E. J. Strait, Phys. Plasmas {\bf 16} 056113 (2009).
\end{thebibliography}

       \end{document}